\documentclass[iop]{emulateapj}

\usepackage{amsmath}    

\usepackage{hyperref}
\usepackage{natbib}
\usepackage{color}
\usepackage{graphicx}
\usepackage{epstopdf}
\usepackage{graphicx}
\received{}
\accepted{}

\slugcomment{to appear in ApJ}
\shorttitle{}
\shortauthors{Wang et al.}

\begin{document}
\title{Detecting Water In the atmosphere of HR 8799 \MakeLowercase{c} \\with $L$-band High Dispersion Spectroscopy Aided By Adaptive Optics}

\author{
Ji Wang\altaffilmark{1,2}, 
Dimitri Mawet\altaffilmark{1},
Jonathan J. Fortney\altaffilmark{3},
Callie Hood\altaffilmark{3}, 
Caroline V. Morley\altaffilmark{4}, and
Bj{\"o}rn Benneke\altaffilmark{5}
} 
\email{ji.wang@caltech.edu}
\altaffiltext{1}{Department of Astronomy, California Institute of Technology, MC 249-17, 1200 E. California Blv, Pasadena, CA 91106 USA}
\altaffiltext{2}{Department of Astronomy, The Ohio State University, 100 W 18th Ave, Columbus, OH 43210 USA}
\altaffiltext{3}{Other Worlds Laboratory, Department of Astronomy \& Astrophysics, University of California, Santa Cruz, CA 95064, USA}
\altaffiltext{4}{Harvard-Smithsonian Center for Astrophysics, Cambridge, MA, USA}
\altaffiltext{5}{Institut de Recherche sur les Exoplan\`etes, Universit\'e de Montr\'eal, Canada}

\begin{abstract}

High dispersion spectroscopy of brown dwarfs and exoplanets enables exciting science cases, e.g., mapping surface inhomogeneity and measuring spin rate. Here, we present $L$ band observations of HR 8799 c using Keck NIRSPEC (R=15,000) in adaptive optics (AO) mode (NIRSPAO). We search for molecular species (H$_2$O and CH$_4$) in the atmosphere of HR 8799 c with a template matching method, which involves cross correlation between reduced spectrum and a template spectrum. We detect H$_2$O but not CH$_4$, which suggests disequilibrium chemistry in the atmosphere of HR 8799 c, and {{this is consistent with previous findings}}. We conduct planet signal injection simulations to estimate the sensitivity of our AO-aided high dispersion spectroscopy observations. We conclude that $10^{-4}$ contrast can be reached in $L$ band. The sensitivity is mainly limited by {{the accuracy of line list used in modeling spectra}} and detector noise. The latter will be alleviated by the NIRSPEC upgrade.    

\end{abstract}


\section{Introduction}
\label{sec:intro}

High dispersion spectroscopy provides a way of resolving spectral lines and extracting rich information therein. By monitoring periodic line profile variation, surface inhomogeneity (e.g., clouds and/or spots) of a nearby brown dwarf (BD), Luhman 16 B, has been retrieved by the Doppler imaging technique~\citep{Crossfield2014}. By comparing line width with instrumental line broadening, the spin rate of directly-imaged exoplanets has been measured~\citep{Snellen2014,Bryan2018}. In addition, high dispersion spectroscopy can significantly increase the sensitivity of a coronagraphic system, enabling detection and characterization of rocky planets in habitable zones~\citep{Sparks2002, Riaud2007, Kawahara2014}. Therefore, high dispersion coronagraphy (HDC) emerges as an active area of study for future ground-based extremely large telescopes and space missions~\citep{Snellen2015, Lovis2016, Wang2017a, Mawet2017, Hoeijmakers2018}.

HR 8799 bcde is the only multi-planet system that has been directly imaged~\citep{Marois2008, Marois2010}. The system has been extensively studied by previous observational campaigns. The astrometry of the four planets are measured to a few milliarcsecond precision~\citep{Konopacky2016, Wertz2017}. The atmospheres of planets in this system have been studied by multi-band photometry and low-resolution spectroscopy~\citep[e.g.,][]{Skemer2014, Bonnefoy2016, Zurlo2016}. As one of the most scientifically intriguing and the most studied system, HR 8799 will be a prime target for future HDC instruments~\citep[e.g.,][]{Mawet2016, Lovis2016}. 

Observing the HR 8799 planetary system with high dispersion spectroscopy aided by adaptive optics (AO) is a major step towards future HDC observations. An AO system significantly reduces the stellar light at the planet location compared to the seeing-limited condition. High dispersion observation aided by AO thus reduces a major noise source: photon noise of contaminating stellar light~\citep{Snellen2015}. \citet{Konopacky2013} conducted AO-aided integral field unit (R$\sim$4000) observations of HR 8799 c using Keck OSIRIS. Using a template matching method, which involves cross-correlating the observed spectrum with a template spectrum for the planet or an individual molecular species, they detected water (H$_2$O) and carbon monoxide (CO) in the atmosphere of HR 8799 c.  A similar study was conducted for HR 8799 b and resulted in simultaneous detection of H$_2$O, CO and methane (CH$_4$) in $K$ band~\citep{Barman2015}.   

In principle, higher spectral resolution would lead to a higher peak in a cross correlation function (CCF), thus increasing detection significance. However, there may be practical limits as light is dispersed onto more pixels, e.g., detector noise. As of today, AO-aided high dispersion spectroscopy of exoplanets has been conducted on a few exoplanets for spin measurements~\citep{Snellen2014,Bryan2018} and molecular detection~\citep[][, and references therein]{Hoeijmakers2018}. 

In this paper, we report high dispersion (R$\sim$15,000) observation of HR 8799 c using Keck NIRSPEC in AO mode. We attempt to (1) characterize its atmospheric chemical composition and (2) understand the fundamental and practical limit in AO-aided high dispersion spectroscopy. This paper will shed light on observations with upcoming instruments in the near future, e.g., the upgraded NIRSPEC~\citep{Martin2014} and CRIRES+~\citep{Follert2014}, and the Keck Planet Imager and Characterizer~\citep[KPIC,][]{Mawet2016}. With the higher sensitivity provided by these instruments, there will be more targets and higher signal to noise ratios (SNR) for studies of BDs and exoplanets. 

The paper is organized as follows. We present the observation in \S \ref{sec:observation}. In \S \ref{sec:data_reduction}, we provide details of the procedure to reduce raw data to wavelength-calibrated spectra. We describe data analyses and the extraction of the planet signal in \S \ref{sec:data_analysis}. Results are reported in \S \ref{sec:result}. We conduct sensitivity analyses in \S \ref{sec:sensitivity_analysis} in order to understand the threshold planet/star flux ratio to which our observation and data reduction are sensitive. Conclusion and discussion are given in \S \ref{sec:summary}. 

\section{Keck NIRSPAO Observation of HR 8799~\lowercase{c}}
\label{sec:observation}

\subsection{Instrument Setup}
We observed HR 8799 c using NIRSPEC in AO mode in $L$ band with the `ML" filter. We selected a slit size of 0.068$^{\prime\prime}$x2.26$^{\prime\prime}$. The first value, 0.068$^{\prime\prime}$, is slit width. The slit width is sampled by 5 pixels on the detector, corresponding to a spectral resolution of R$\sim$15,000. We chose this slit width to ensure sufficient planet flux enters the slit in the presence of guiding error and the resulting flux loss. The second value, 2.26$^{\prime\prime}$, is the slit length. The slit length ensures that both the star and planet c are contained within the slit. {{Planet-star separation and position angle were calculated to milliarcsec and a tenth of a degree precision for the epochs in which observations were conducted. Two independent codes for astrometry prediction were used~\citep{Wang2016, Wertz2017} and they gave consistent values for separation and position angle (see Table \ref{tab:obs_summary}). }} 

We set slit angle to match with the position angle of HR 8799 c. In this set up, spectra of both star and planet are recorded on detector. We use the stellar spectrum as a reference for both telluric line absorption and planet position. HR 8799 is an early type star (T$_{\rm{eff}}=7435$ K and $\log\ g=4.35$) and rotates relatively fast with a v$\sin i$ of 37.5 $\rm{km}\cdot\rm{s}^{-1}$~\citep{Gray1999}, so we used it as a telluric standard. Because we know precisely the planet-star separation, the planet spectrum, albeit not visible in raw data, can be traced relative to the stellar spectrum. 
 
We observed astrometric standard stars to measure the plate scale (in the unit of arcsec per pixel) along the slit direction. This is used to convert the planet-star separation in the unit of arcsec to pixels along slit. HO 482 AB~\citep{Prieur2014} was chosen from the Sixth Catalog of Orbits of Visual Binary Stars~\citep{Hartkopf2001}\footnote{http://ad.usno.navy.mil/wds/orb6/orb6frames.html}. At the time of observation (UT 2016 Aug 14), the position angle and angular separation were 15.10 degree and 0.558$^{\prime\prime}$ for HO 482 AB. We reduced the astrometric standard stars the same way as we did for HR 8799 c data (see \S \ref{sec:data_reduction}). The resulting plate scale was $0.0179\pm0.0006^{\prime\prime}$. {{The planet scale uncertainty was calculated using the Root-Mean-Square (RMS) of the measurements from 5 spectral orders. The actual uncertainty of plate scale may be larger because of systematic errors in determining the orbit of the astrometric standard star, HO 482 AB. According to ~\citet{Prieur2014}, we estimated the actual uncertainty of plate scale to be 0.001$^{\prime\prime}$.}}


\subsection{Observing in $L$ Band}
We observed HR 8799 c in $L$ band on UT 2016 Aug 12-14, 2017 Jul 6, and 2017 Nov 6. Data obtained on Aug 14 were not used because of one-magnitude cloud extinction and highly variable sky background and water content. A summary of observations is given in Table \ref{tab:obs_summary}. 

We used the ``Stare" mode in which a target stays at a fixed position in the slit without any dithering pattern. The ``Stare" mode allowed for a more effective duty circle as no overhead was incurred by dithering. Total on-target time was 15.6 hours. Compared to the wall time duration of 18.2 hours, the duty circle was 86\%.

Exposure time was set to be 60 second and 3 coadd per frame. We obtained 74, 79, 78, and 81 frames on four half nights, respectively. On UT 2016 Aug 12-13 and 2017 Jul 6, peak flux recorded on the detector was $\sim$8,000 ADU depending on target airmass and seeing condition. On 2017 Nov 6, peak flux was $\sim$4,000 ADU because of poor AO performance despite good seeing condition.

{{Sky background in $L$ band is a major noise source. We therefore list in Table \ref{tab:obs_summary} the range of sky background fluctuation, which was mostly between $\sim$1,000 and 2000 ADU. The values were well below the non-linear range for NIRSPEC detector (15,500 ADU) and the charge persistence threshold (4000 ADU)\footnote{https://www2.keck.hawaii.edu/inst/nirspec/Specifications.html}.
}}

\subsection{Observing in $K$ Band}
We observed HR 8799 c in $K$ band on UT 2016 Aug 11. The $K$ band data allowed us to independently measure the absolute radial velocity (RV) for HR 8799 (\S \ref{sec:st_rv}). Exposure time was set to be 60 second and 1 coadd per frame. Peak flux recorded on the detector is $\sim$10,000-20,000 ADU (gain = 5.8 $e^{-1}$ per ADU) depending on target airmass and seeing condition.




\section{Data Reduction}
\label{sec:data_reduction}

\subsection{Adapting PyNIRSPEC to ``Stare"-Mode Observation}
\label{sec:pynirspec_stare}
We reduced raw data from NIRSEPC using the Python-based package PyNIRSPEC~\citep{Boogert2002, Piskorz2016,Bryan2018}. Since PyNIRSPEC is customized for data obtained with a dithering pattern, we adapted the code so that it worked for the ``Stare" mode observation. Specifically, we treated HR 8799 data as the data for dither position ``A". We added another set of mock-up data with zero values for all pixels and treated the mock-up data set as the data for dither position ``B". In this way, PyNIRSPEC can successfully process the data obtained in the ``Stare" mode. 

\subsection{Data Reduction Procedures in PyNIRSPEC}
Raw images were subtracted by darks and then flat fielded. Bad pixels were identified in dark frames and their values were replaced by interpolating surrounding pixels.  

\begin{figure*}
\epsscale{1.0}
\plotone{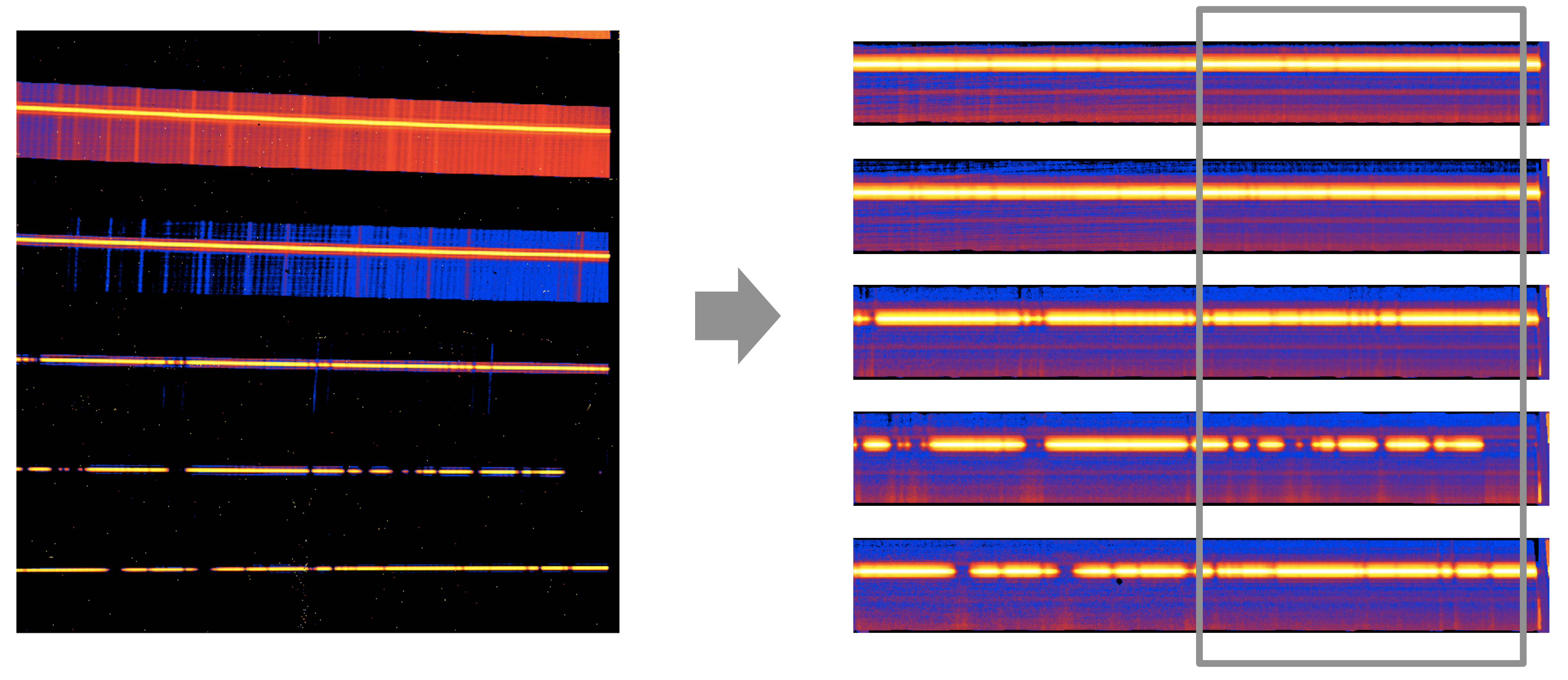}
\caption{Reducing $L$-band raw data (left) into rectified 2-d spectra (right). On the left side of the detector, the noise pattern is repeated every 8th row. In raw data, the height of each order is 130 pixels on the left side and 120 pixels on the right side. The spectral trace of HR 8799 is clearly visible along with sky emission lines. In contrast, the spectral trace of HR 8799 c is buried in noise and thus not visible from either raw data or reduced 2-d spectra. The grey box indicates the region used for subsequent data analyses.  
\label{fig:HR8799c_L_raw_order}}
\end{figure*}

The raw images were then divided into different orders (Fig. \ref{fig:HR8799c_L_raw_order}). Each order was processed independently including the following procedures: rectification and wavelength calibration (as detailed in \S \ref{app:dr_details}). The final data products of PyNIRSPEC are wavelength-calibrated rectified 2-d spectra. 

\section{Data Analyses}
\label{sec:data_analysis}
HR 8799 c is $2\times10^{-4}$ times the flux of its host star in $L$ band~\citep{Currie2014}. In a reduced 2-d spectrum, the planet signal is entirely overwhelmed by the sky background emission. Stellar continuum has a flux level of a few tens of thousands electrons per pixel. In comparison, the planet signal is only a few electrons per pixel. Additionally, correlated noise from the detector is another form of non-negligible noise and needs to be removed. We describe our strategy of cleaning the planet spectrum based on principle component analysis (PCA) in \S \ref{sec:pca}. In \S \ref{sec:template}, we describe the procedures of extracting planet signal using a template matching method. 

\subsection{Grouping 2-d Spectra}
\label{sec:group_spec}
In principle, one could combine all 2-d spectra and conduct data analyses to extract planet signal. However, we found that features in the 2-d spectra changed over time. For example, speckles appeared and disappeared in slit. The potential for capturing time-varying features in spectra is lost when stacking up all data. 

On the contrary, using the 2-d spectrum from a single frame is prone to stochastic phenomena such as cosmic ray events. {{After experimenting a set of grouping numbers ranging from 6 to 40 frames, we decided to group every 10 frames and use the median for subsequent analyses.}} This corresponds to a 34-minute time duration for $L$-band observation. We found that in such a time scale, speckle features in the obtained 2-d spectra remained relatively stable and cosmic ray events were removed reasonably well.  

\subsection{Principle Component Analysis}
\label{sec:pca}
Planet signal is buried in a variety of noise sources including photon noise, speckle noise and detector noise. Among these noise sources, photon noise is fundamental and impossible to circumvent. However, speckle noise and detector correlated noise may be removed. This can be done by building a model of patterns caused by these noises at positions excluding the planet location. The model is then applied to the planet location to correct for the noise patterns. This goal can be achieved by PCA which is detailed below. 

\subsubsection{Extracting 1-d Spectrum}
We describe as follows the procedure to extract a 1-d spectrum from a 2-d rectified wavelength calibrated spectrum. The slit direction of the 2-d spectrum was oversampled by a factor of $\sim$4 with a plate scale of 0.0045$^{\prime\prime}$ per pixel. We extracted a 1-d spectrum at a given slit location with the following equation:
\begin{equation}
\label{eq:extraction}
{\rm{S}}_i = {\rm{I}}_{i, j}\cdot{\rm{K}}_j(\delta),
\end{equation}
where $i$ and $j$ are pixel numbers along the dispersion direction and the slit direction, ${\rm{S}}_i$ is the 1-d spectrum at the $j$th pixel along the slit, ${\rm{I}}_{i, j}$ is a 2-d spectrum, and ${\rm{K}}_j$ is a kernel function in the form of a Gaussian function that centers at the $j$th pixel with a varying $\delta$, i.e., the standard deviation of the Gaussian function. The chosen $\delta$ ranged from 0.009$^{\prime\prime}$ to 0.063$^{\prime\prime}$, i.e., from a fraction of the stellar point spread function (PSF) full width at half maximum (FWHM) to about one PSF FWHM. Wavelength for each pixel $i$ can be calculated using the wavelength solution obtained in \S \ref{sec:wav_sol}. 

\subsubsection{Building a Reference Library for 1-d Spectra}
Correlated noise may exist in both spatial and temporal dimension. In order to capture the correlated noise and remove it from the planet spectrum, we built a 1-d spectral library that encompassed both spatial and temporal dimension (see an illustration in top-left panel of Fig. \ref{fig:illus_L}). 
For each 2-d spectrum, we extracted 1-d spectra at different positions along the slit direction. The locations for the 1-d spectra excluded two ends of the slit with a height of $2\times\delta$ and the areas in the vicinity of the planet, $\pm3\times\delta$ from the planet location. The 1-d spectra were extracted with an increment of $2\times\delta$ along the slit. 

We repeated the same procedure for 2-d spectra taken at different times. This allowed us to build a 1-d spectral library for the following PCA. 

\subsubsection{PCA For Highly Correlated Spectra}
Not all spectra in the library were useful in identifying correlated noise in 1-d planet spectrum. Intuitively, only those spectra that were close to the planet spectrum in spatial and temporal dimension were correlated with the planet spectrum. We identified the highly correlated spectra in library with the cross correlation method. We cross-correlated the planet spectrum with all spectra in the library and picked the ones that gave a significant peak (above 3 times of root mean square, i.e., RMS) in the cross correlation function (CCF). Typically, we had at least 30 highly correlated spectra for PCA (see an illustration in the top-right panel of Fig. \ref{fig:illus_L}). 

{{We subtracted the median value off the highly correlated spectra and normalized each spectrum by dividing the difference of the 90 and 10 percentile values.}} Then we conducted PCA and found the principle components in these spectra, which were later subtracted from the planet spectrum. The first few principle components are normally caused by speckle noise, detector correlated noise, and telluric features. We describe the math of PCA with the following equation:
\begin{equation}
\label{eq:pca1}
\rm{S} = \rm{U}\rm{E}\rm{V}^{\rm{T}},
\end{equation}
where $\rm{S}$ is an $n\times m$ array, $m$ is the number of pixels of each spectrum and $n$ is the number of spectra that are highly correlated with the planet spectrum, $\rm{U}$, $\rm{E}$ and $\rm{V}^{\rm{T}}$ are the result of single value decomposition (SVD) of S. $\rm{U}$, an $n\times n$ array contains the left singular vectors, $\rm{E}$ is an $n\times n$ diagonal matrix of the singular values (or the eigenvalues), and $\rm{V}^{\rm{T}}$, an $n\times m$ array, contains the right singular vectors, where ${\rm{T}}$ denotes the transpose operation. Columns in $\rm{V}$ are eigenvectors, representing principle directions.   

\subsubsection{Removing Principle Components From Planet Spectrum}
We projected the planet spectrum onto principle directions and subtracted the first few projections from the planet spectrum. This helped to remove any correlated noise from the planet spectrum. The operation is described by the following equation:
\begin{equation}
\label{eq:pca}
{\rm{\tilde{P}}}_i = {\rm{P}}_i - \sum_{k=1}^{N}({\rm{P}}_i\cdot{\rm{V}}_k)\times{\rm{V}}_k,
\end{equation}
where ${\rm{P}}$ and $\rm{\tilde{P}}$ are the planet spectra before and after the first $N$ principle components are removed, $i$ is the subscript for pixel number, $k$ is the subscript for eigenvector number, ${\rm{V}}_k$ is the $k$th eigenvector from PCA. 



\subsubsection{Combining Planet Spectra at Different Times}
We denote the planet spectrum after PCA as the reduced planet spectrum. Since the reduced planet spectra were obtained at different times, we describe how they were combined to form a final planet spectrum for subsequent analyses.

\begin{figure*}
\epsscale{1.1}
\plotone{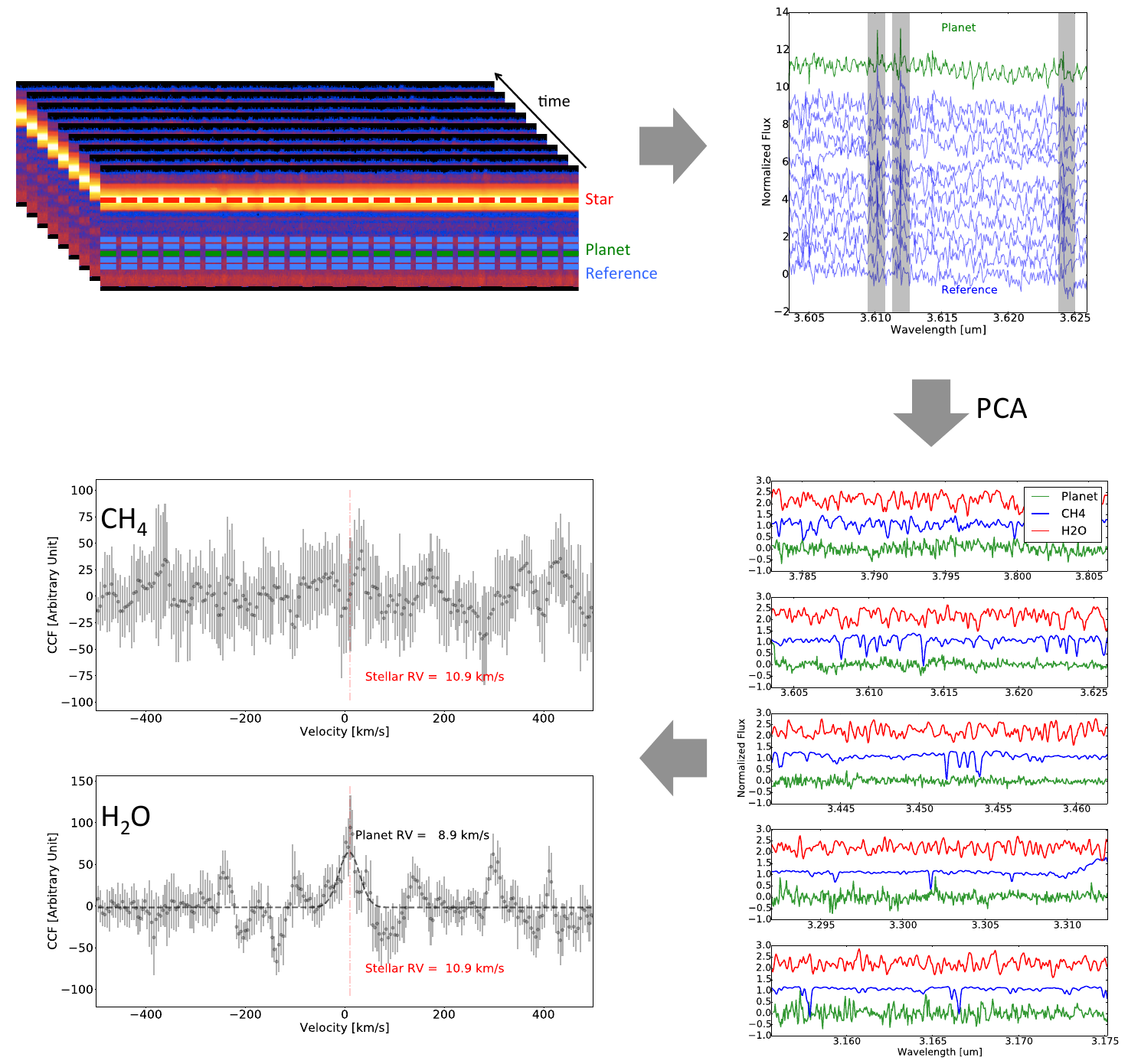}
\caption{Illustration of data reduction and analyses. Top left: building a spectral reference library. Star, planet, and reference traces are marked in red, green and blue, respectively. Top right: raw planet spectrum (green) vs. reference spectra (blue). The spectra share many similar features, e.g., regions marked in grey. Bottom right: PCA removes common features shared by raw planet spectrum and reference spectra. Cleaned planet spectrum (green) is cross-correlated with template spectra (CH$_4$ - blue, H$_2$O - red). Bottom left: CCFs and stellar RV marked as red dashed lines. 
\label{fig:illus_L}}
\end{figure*}

We subtracted the median value from each reduced planet spectrum. We formed a $p\times m$ array of reduced planet spectra for PCA, where $m$ is the number of pixels in each reduced planet spectrum and $p$ is the number of spectra at different times. We used the first principle component of the $p\times m$ array, i.e., $\rm{E}_{11}\times \rm{V}_1$ (see Equation \ref{eq:pca}), as the final planet spectrum. This is very similar to taking the median of all reduced planet planet spectra. For example, the $L$ band final planet spectrum is shown on the bottom-right panel of Fig. \ref{fig:illus_L}. 

\subsection{Template Matching}
\label{sec:template}

\subsubsection{Generating Templates}
We generated high-resolution emission spectra using well-tested atmosphere modeling tools.  Temperature structures were generated for models that iterated to radiative-convective equilibrium, in the absence of a parent star, assuming equilibrium chemistry, following \citet{Marley2002,Fortney2008,Marley2012}.  We then generated line-by-line spectra at $R=200,000$ using the code described in the Appendix of \cite{Morley2015}.  To isolate the contributions of particular molecules, most spectra were generated with one or two molecules, namely CH$_4$ and H$_2$O, with all other opacity removed.  The model opacity database was described in detail in \citet{Freedman2008}, with updates in \citet{Freedman2014} and references therein.  In particular, for CH$_4$ and H$_2$O respectively, the first-principles line lists of \cite{Yurchenko2014} and \citet{Barber2006} were used.

\subsubsection{Template Matching With Cross Correlation}
Planet signal can be extracted by cross correlating the final reduced planet spectrum with a template spectrum. While the planet signal is overwhelmed by noise at single pixel level, the cross correlation operation integrates planet signal of all pixels with a matching template. This technique has proven to be effective in extracting faint planet signal in the presence of strong contamination~\citep[e.g., ][]{Snellen2010, Piskorz2016}. 

\subsubsection{Combining Cross Correlation Functions}
Since cross correlation was performed for each spectral order, we describe here how CCFs were combined to construct the final CCF. Each CCF was normalized by its RMS value and resampled into the same velocity scale. Next, CCFs for different orders were summed up to form a CCF for a single night. {{Note that we gave equal weight to each order in the summation process. While the weight may be different and determined by CCF SNR for each order~\citep{Bouchy2001}, we were concerned whether the measured CCF SNR for each order accurately reflected the information content therein given unknown noise properties and the choice of window function. Alternatively, we could in principle inject simulated signal in each order and estimate the contribution to combined CCF. However, model uncertainties such as molecular abundance, mixing ratio, and cloud coverage may significantly alter the flux of the planet and the spectral information in each order, making the determination of order contribution impractical. As such, we assigned equal weight to each spectral order to construct a combined CCF for each night. }}

To construct the final CCF, we also summed up CCFs for different nights with weights that are determined by injection simulation (see \S \ref{sec:injection_reduced}). The weights are a measure of data quality for each night. The final CCF was used to detect atmospheric molecular species and to access the detection significance. 

\subsubsection{Using Templates For Single Molecular Species}
{{Templates involving multiple molecular species are more uncertain than templates for single molecular species because the former has an additional parameter, i.e., mixing ratio between different species (as will be shown in \S \ref{sec:implications})}}. Therefore, we use template spectrum for single molecular species, namely H$_2$O and CH$_4$. These two species dominate opacity in $L$ band. Using templates for single molecular species circumvents the caveat of miscalculating the relative abundance. The results for single species can be combined to create a stronger peak in the resulting CCF~\citep{Konopacky2013, Hoeijmakers2018} if the model for multi-species is correct. 

\subsubsection{Exploring Optimal Parameters For Planet/Molecular Detection}
In the data analyses procedures described in this section, there are several parameters that can be changed to optimize the data analyses. These parameters are (1) the standard deviation of the Gaussian function used to extract the 1-d spectrum, which is referred to as extraction width; (2), the number of principle components to be removed from the planet spectrum; and (3), the planet location. The planet location may vary in pixel space because of the uncertainty of the plate scale we measure. We stepped through the above three parameters within a reasonable range in order to fully explore the optimal parameters in data analyses. Specifically,  extraction width ranged from 0.009$^{\prime\prime}$ to 0.063$^{\prime\prime}$; a maximum number of 20 was set for principle components to be removed; positional error for the planet was assumed to be 4\%, consistent with the measurement error of the plate scale. The optimization was done on a nightly basis because observation condition varied from night to night.  

\section{Result}
\label{sec:result}
We find a peak in the H$_2$O CCF that shows a consistent RV with the stellar RV of HR 8799 (see the bottom-left panel of Fig. \ref{fig:illus_L}). We find no significant peak in the CH$_4$ CCF.

\subsection{Detection of H$_2$O in $L$ band}
We detect H$_2$O in $L$ band in the atmosphere of HR 8799 c with a SNR of 4.6 (Fig. \ref{fig:illus_L}). CCF SNR is defined as the ratio between the peak value and the CCF RMS. Below we provide multiple lines of evidence for the detection. 

\subsubsection{Uncertainty of Each CCF Data Point}
Each data point of the CCF has its own uncertainty that is estimated using the jackknife resampling method. Specifically, we create 8 subsamples by each time removing one of the 8 grouped $L$ band spectra in each night. We then redo the data analyses on the 8 subsamples, resuting in 8 CCFs. An estimate of each data point of the CCF $\bar{x}$ is:
\begin{equation}
\label{eq:jn_mean}
\bar{x}=\frac{1}{n}\sum_{i=1}^{n}\bar{x_i},
\end{equation}
where $\bar{x_i}$ is the CCF data point of the subsample that leaves the $i$th subsample out and $n$ is the total number of subsamples. The variance of the CCF data point is:
\begin{equation}
\label{eq:jn_std}
Var=\frac{n-1}{n}\sum_{i=1}^{n}(\bar{x_i}-\bar{x})^2.
\end{equation}
We use 1-$\sigma$ value as the uncertainty of each data point of the CCF, i.e., $\sigma=\sqrt{Var}$.

\subsubsection{Consistent Planet and Star Radial Velocities}
\label{sec:st_rv}
The H$_2$O CCF peak has a RV of $8.9\pm2.5$ km$\cdot\rm{s}^{-1}$. Positive RV indicates blue shift throughout this paper. In comparison, we measure the stellar RV at $10.9\pm0.5$ km$\cdot\rm{s}^{-1}$. We describe below how planet RV and stellar RV are calculated.

We fit the CCF with a Gaussian function. The peak of the Gaussian function is $8.9\pm2.5$ km$\cdot\rm{s}^{-1}$. We estimate the uncertainty by randomizing CCF data points based on their uncertainties and repeating the fitting process. 

We measure stellar RV using Hydrogen Bracket gamma line in $K$ band ($\lambda_0=2.16612\ \mu$m). There are two telluric lines next to the Bracket gamma line at $\lambda=2.16345\ \mu$m and $\lambda=2.16869\ \mu$m. We can therefore calibrate the wavelength in the Bracket gamma line region. Anchoring the two telluric lines, we use a linear fit to calculate wavelength as a function of pixels. We then use a quadratic function to fit for the line center of the Bracket gamma line. We measure the stellar RV at $10.9\pm0.5$ km$\cdot\rm{s}^{-1}$. In comparison, the RV for HR 8799 is $12.6\pm1.4$ km$\cdot\rm{s}^{-1}$~\citep{Gontcharov2006}. These two values are consistent within 1-$\sigma$, and both values agree with the planet RV. 

Planet RV due to orbital motion needs to be taken into consideration when comparing planet RV and stellar RV. Semi-amplitude of planet orbital RV is estimated at 2.7 km$\cdot\rm{s}^{-1}$ assuming a nominal plant mass of 7 Jupiter mass, a nominal stellar mass of 1.47 solar mass and orbital parameters reported in~\citet{Wertz2017}. The time of periastron passage for HR 8799 c is largely uncertain because only partial orbit has been observed. Planet orbital RV can range from -2.7 km$\cdot\rm{s}^{-1}$ to 2.7 km$\cdot\rm{s}^{-1}$ depending on the value of the time of periastron passage within measurement uncertainty. However, the measured CCF RV is not significantly altered by the planet orbital RV even when the full range of possible planet orbital RV is considered. 

\subsubsection{Low False Positive Rate}
{{CCF SNR is sensitive to the choice of window function and noise properties, we discuss here a more appropriate way of assigning detection significance. We investigate the probability of the CCF peak arising from random fluctuation. We follow the same data analyses procedure except that we permutate the final spectra before cross-correlating with a template spectrum. We repeat this exercise for $10^5$ times. We find that 2 of the randomization processes produce a CCF peak with a SNR higher than 4.6. This corresponds to a false positive rate of $2.0\times10^{-5}$, which translates into a detection significance of 4.2-$\sigma$.}}

\subsubsection{Favorable Bayesian Information Criterion}
We compare the Bayesian information criterion (BIC) for two models, a Gaussian function and a flat line. This is to test if a Gaussian fit to the CCF has a lower BIC than a flat line fit, i.e., the CCF is better represented by a Gaussian function than a flat line. We calculate BIC with the following equation:
\begin{equation}
\label{eq:bic}
BIC=\ln(n)\cdot k - 2\ln(\hat{L}),
\end{equation}
where $n$ is the number of data points of CCF, $k$ is the number of free parameters in a model. For the Gaussian model we use, there are 4 free parameters including a y offset from zero, amplitude, mean and standard deviation. The flat line model has only one free parameter, i.e., the y offset from zero. In Equation \ref{eq:bic}, $\ln(\hat{L})$ is the log-likelihood function:
\begin{equation}
\label{eq:log_likelihood}
\ln(\hat{L})=\sum_{i=1}^{n}\left(-\frac{1}{2}\ln(2\pi)-\frac{1}{2}\ln(\sigma_i^2)-\frac{(x_i-\mu)^2}{2\sigma_i^2}\right),
\end{equation}
where $x_i$ is the $i$th data point of a CCF, $\sigma_i$ is the associated measurement uncertainty, and $\mu$ is the predicted value at the $i$th data point by a model. 

Based on Equation \ref{eq:bic} and \ref{eq:log_likelihood}, the difference of BIC between a Guassian model and a flat line model is -49.9, strongly favoring a Guassian model. 

\begin{figure}
  \centering
  \begin{tabular}[b]{c}
    \includegraphics[width=.95\linewidth]{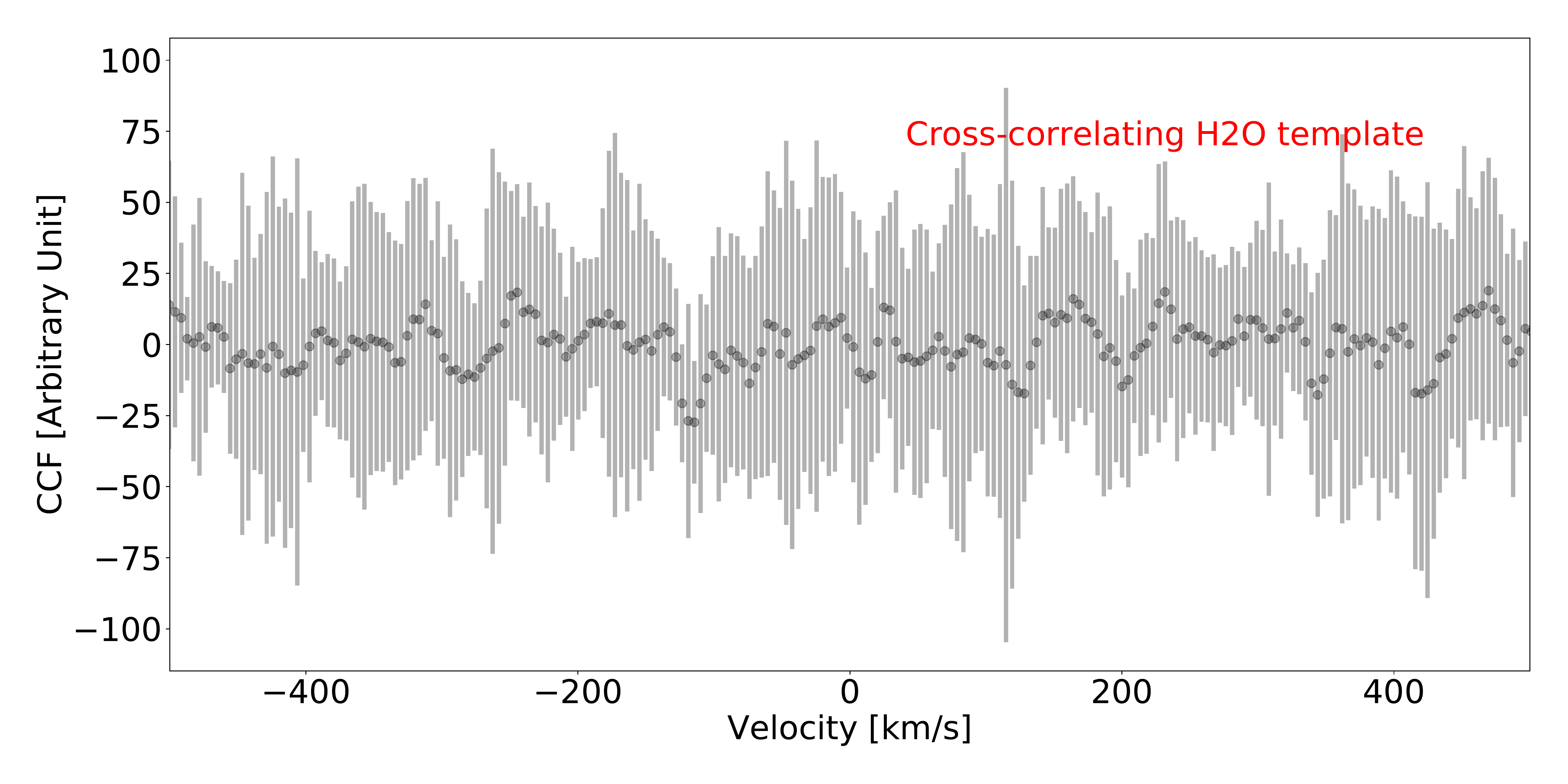}  \\
    \includegraphics[width=.95\linewidth]{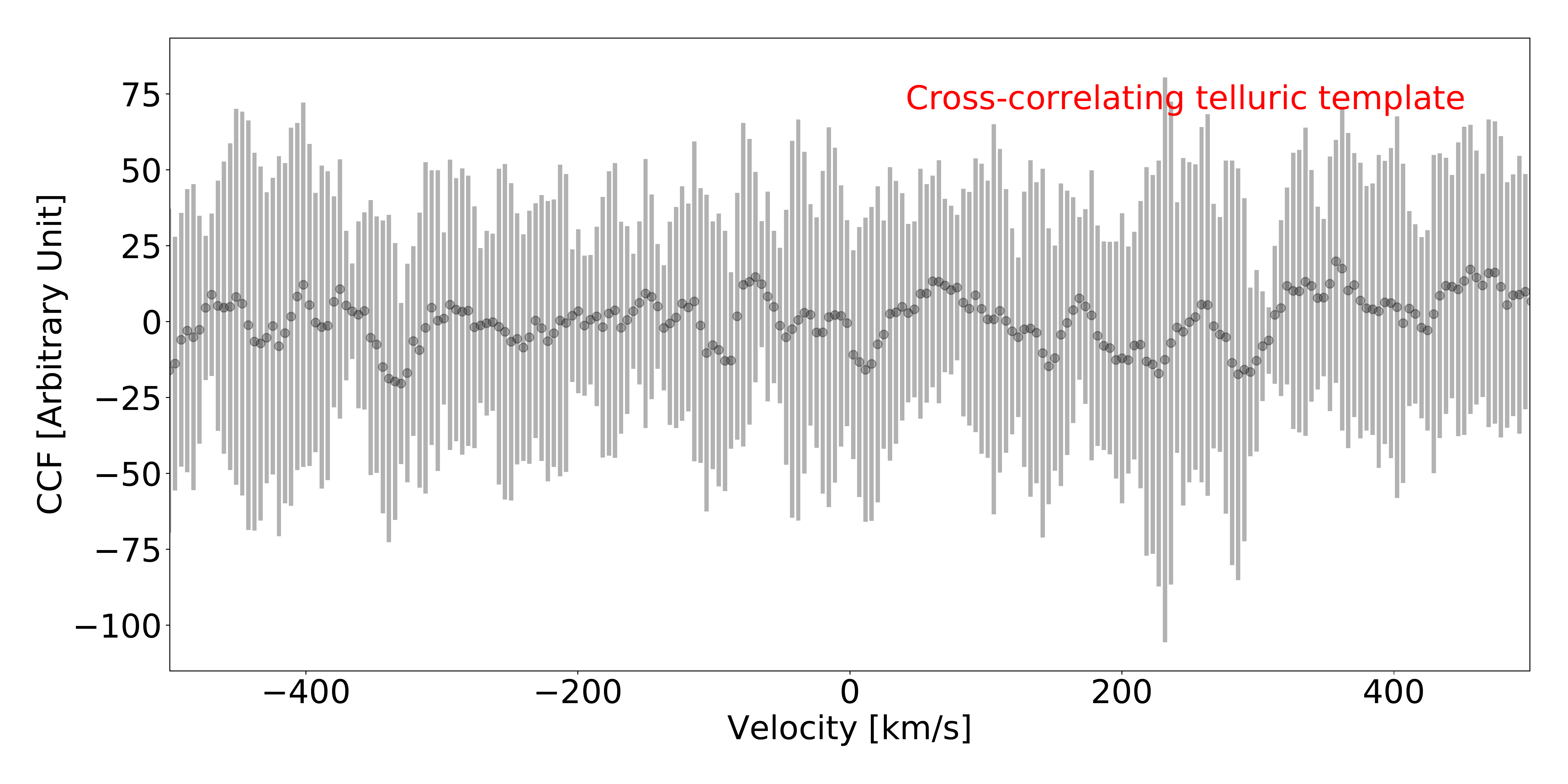}  \\
    \includegraphics[width=.95\linewidth]{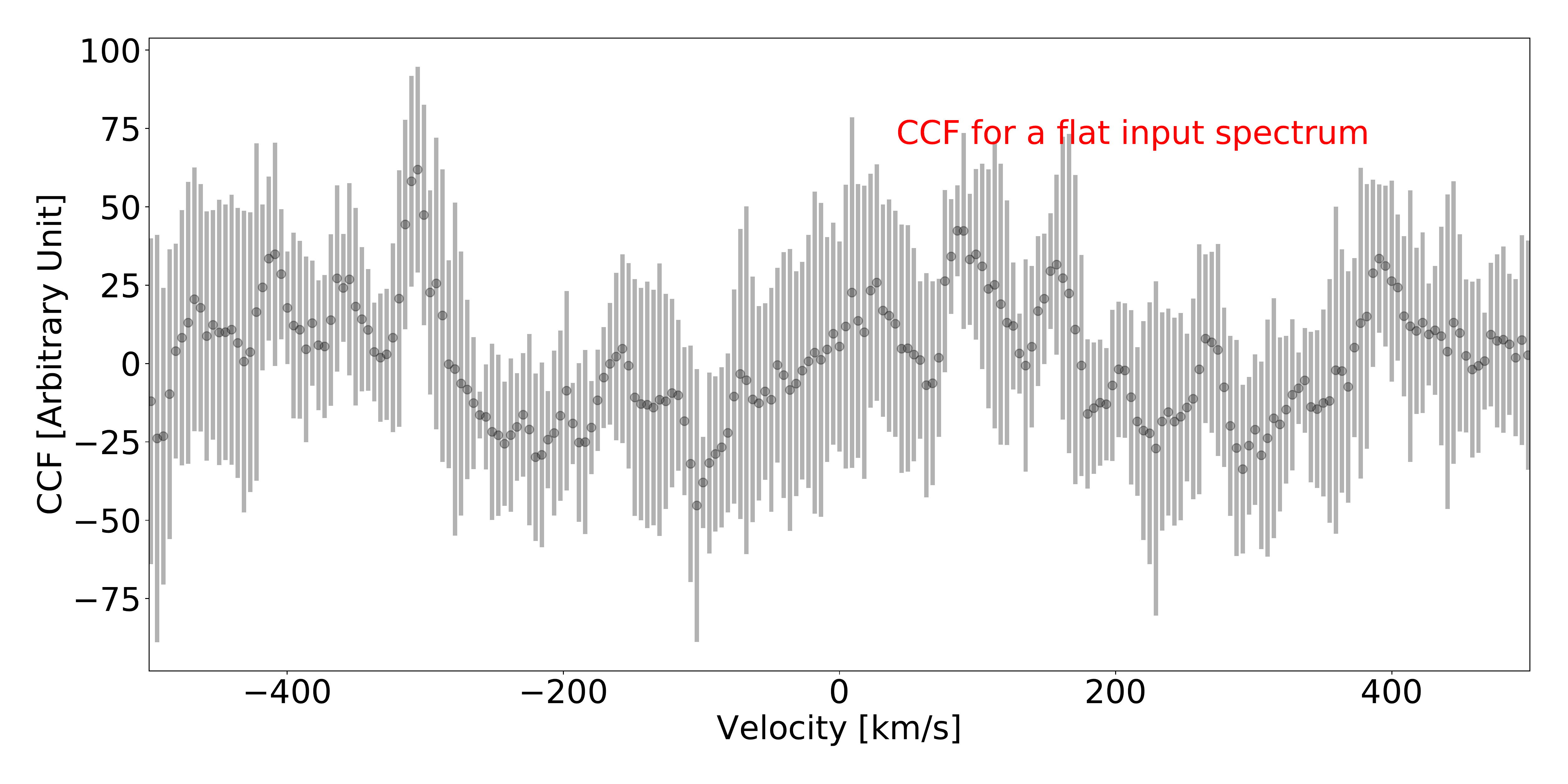}  \\

  \end{tabular} \qquad
  \caption{Top: CCF between the stellar spectrum (after removing telluric lines and correlated noise with PCA) and a H$_2$O template spectrum. This demonstrates that the PCA residual does not generate a peak at the star location. Middle: same as top except that the template spectrum is now a telluric line spectrum. This demonstrates that telluric lines are successfully removed by PCA and does not generate a peak at the star location. Bottom: CCF for a flat (featureless) injected spectrum at a non-planet location. This demonstrates that there is no peak in CCF even if there is a planet signal but with a flat spectrum. \label{fig:test}}
\end{figure} 

\subsubsection{Non-detection of H$_2$O at the Star Location}

{{Since the Earth atmosphere also contains the molecular species that we are searching for, we test here if sky, telluric, and other correlated noise removal residual would result in a false detection. We change the search location to the star location and perform the data analyses as described in \S \ref{sec:data_analysis}. To demonstrate that the PCA residual (after removing telluric lines and correlated noise) does not cause a false detection, we cross-correlate the residual with a H$_2$O template. As shown in the top panel of Fig. \ref{fig:test}, no significant CCF peak is found. 

We also demonstrate that telluric lines are successfully removed. This is shown by cross-correlating the PCA residual with a telluric line template. The resulting CCF does not show significant peak (middle panel of Fig. \ref{fig:test}).
}}

\subsubsection{Non-detection of H$_2$O When Injecting a Flat Spectrum}
{{To further test the validity of the H$_2$O detection as shown in Fig. \ref{fig:illus_L}, we show here the CCF peak only appears when there is a planet signal and when the spectrum of the planet matches with the template. As will be described in \S \ref{sec:injection_reduced}, we inject a flat (featureless) spectrum with a planet-star contrast of $2\times10^{-4}$ at a non-planet location (sep=1.05$^{\prime\prime}$). After applying the same data analysis procedure as the real data, there is no significant CCF peak (bottom panel of Fig. \ref{fig:test}).  

}}

\subsection{Non-detection of CH$_4$}

The CCF for CH$_4$ is consistent with flat within error bars (see Fig. \ref{fig:illus_L}). CH$_4$ is therefore not detected with significance in $L$ band data. More discussions are given in \S \ref{sec:implications}.

\section{Sensitivity Analysis}
\label{sec:sensitivity_analysis}
In this section, we attempt to understand the fundamental sensitivity of the observation of HR 8799 c and the level of planet signal loss at each stage of data reduction and analysis. We start the sensitivity analyses from a simulation that considers a realistic noise budget (\S \ref{sec:simul}). This allows us to understand the fundamental sensitivity of the observation in the presence of a variety of noise sources.

We then inject planet signal in the reduced spectra (\S \ref{sec:injection_reduced}) and in the raw data (\S \ref{sec:injection_raw}). By comparing the results of the injection experiment and the simulation in \S \ref{sec:simul}, we can identify the stage where the planet signal is significantly compromised. This diagnosis points to directions for future improvement. 
\subsection{Simulation With Realistic Noise Budget}
\label{sec:simul}
\subsubsection{Introduction of Simulation Procedures}
The details of the simulation are provided in Section 2 in~\citet{Wang2017a}. We briefly describe the procedures in the simulation. The flux recorded on a detector can be described by the following equation:
\begin{equation}
\label{eq:f_observed}
\rm{f_{\rm{detector}}} = (f_{\rm{planet}}+f_{\rm{star}}\times C)\times f_{\rm{transmission}} + f_{\rm{sky}},
\end{equation}
where $C$ is the starlight reduction factor at the planet location. 

Noise is then added to the flux with the following equation:
\begin{equation}
\label{eq:f_noise}
noise = \sqrt{\rm{f} + \rm{n}_{\rm{exp}} \times RN^{2} + \rm{dark}\times t_{\rm{exp}} },
\end{equation}
where f is the flux incident on the detector, followed by terms for readout noise (RN) and dark current (dark), where $\rm{n}_{\rm{exp}}$ is the number of readout within a total observation time $t_{\rm{exp}}$.
Parameters used in the simulation are provided in Table \ref{tab:telescope_instrument} and Table \ref{tab:HR8799c}. 

The simulated spectrum is then passed to a pseudo data reduction pipeline that removes sky emission and telluric lines, which results in a reduced planet spectrum. The reduced planet spectrum is cross-correlated with a template spectrum. The resulting CCF is used for planet detection and assessment of detection significance. 
We simulate 100 observations and record the median of CCF SNRs.  

\subsubsection{Spectra Used in Simulation}
We use Fortney-group model spectra as the input planetary spectra. For the stellar spectrum, we use a BT-Settl spectrum~\citep{Baraffe2015} with $T_{\rm{eff}} = 7400$~K and log(g) = 4.5. The metallicity [Fe/H] is set to zero for both planet and star. Planet and star fluxes are adjusted such that the model flux is consistent with the absolute flux measured from photometry. 

We consider two scenarios. In the first scenario, the input spectrum and the template do not match. We use a planet model with 1\% CH$_4$ mixing ratio (with respect to equilibrium CH$_4$ abundance) as the input spectrum, and a planet model with 100\% CH$_4$ mixing ratio (i.e., the equilibrium CH$_4$ abundance) as the template spectrum. In the second scenario, the input spectrum matches with the template spectrum, where both spectra are assumed to be the planet model spectrum with 100\% CH$_4$ mixing ratio. 




\subsubsection{Simulation Results}

In the mismatch scenario, the input spectrum is essentially dominated by H$_2$O opacity because of a low CH$_4$ mixing ratio. Therefore, H$_2$O is detected at a much higher significance than CH$_4$ (15.65 vs. 3.59, see also Table \ref{tab:simul_result}). If any unknown noise is unaccounted for in our simulation, CH$_4$ may be undetectable and H$_2$O would be detected at a lower significance. This is consistent with our observational results, i.e., detection of H$_2$O at 4.6-$\sigma$ and non-detection of CH$_4$.

The mismatch scenario has two implications. First, it is better to use a template spectrum consisting of only one molecular species for cross correlation if mixing ratio of different species is uncertain. Using a H$_2$O template spectrum results in a 15.65-$\sigma$ detection whereas using a template spectrum with a mismatched mixing ratio (1\% vs. 100\% CH$_4$ mixing ratio) results in a much lower detection significance (4.96-$\sigma$). 

Second, a mismatched template spectrum could lead to a much reduced detection significance. In this case, the detection significance is reduced by at least a factor of 3. This stresses the importance of a matched template, or alternatively, the potential of using the template matching technique for spectral inference. That is, we can explore a large parameter space to generate model spectra. The best matched model spectra (i.e. the ones that give the highest detection significance) would help to infer the physical and chemical conditions on an exoplanet. 

In the match scenario, only CH$_4$ is detected because CH$_4$ dominates the opacity if the atmosphere is in chemical equilibrium. This is at odds with our observation. More discussion on chemical equilibrium, CH$_4$ mixing ratio, and modeling uncertainty will be given in \S \ref{sec:summary}.

\subsection{Injecting Planet Signal In the Reduced Planet Spectra}
\label{sec:injection_reduced}
We inject simulated planet signal into reduced data and test if HR 8799 c can be detected. 

\subsubsection{Creating Injection Signal}
\label{sec:create_injection}
We use an $L$ band planetary model spectrum containing H$_2$O and CH$_4$ as the simulated planet spectrum. The planet spectrum is multiplied with the Earth's telluric spectrum and then spectrally blurred to match the spectral resolution in observation (R=$15,000$). The spectrum is normalized by dividing by the median value. 

\subsubsection{Injecting Into Reduced Spectra}
We calculate the stellar PSF along the slit by taking the median value of each row. We obtain the integrated stellar flux per spectral channel by integrating flux along the slit at the position of the stellar PSF. Planet PSF is assumed to be the same as the stellar PSF but with much reduced flux and a positional shift along the slit. To inject simulate planet signal for each spectral channel, we (1) multiply the stellar PSF by a planet-star contrast to form a planet PSF; (2) shift the planet PSF to the planet location; and (3) multiply the planet PSF by a value that corresponds to the normalized planet spectrum for that spectral channel. 

\subsubsection{Injection Results}
\begin{figure*}
\epsscale{1.0}
\plotone{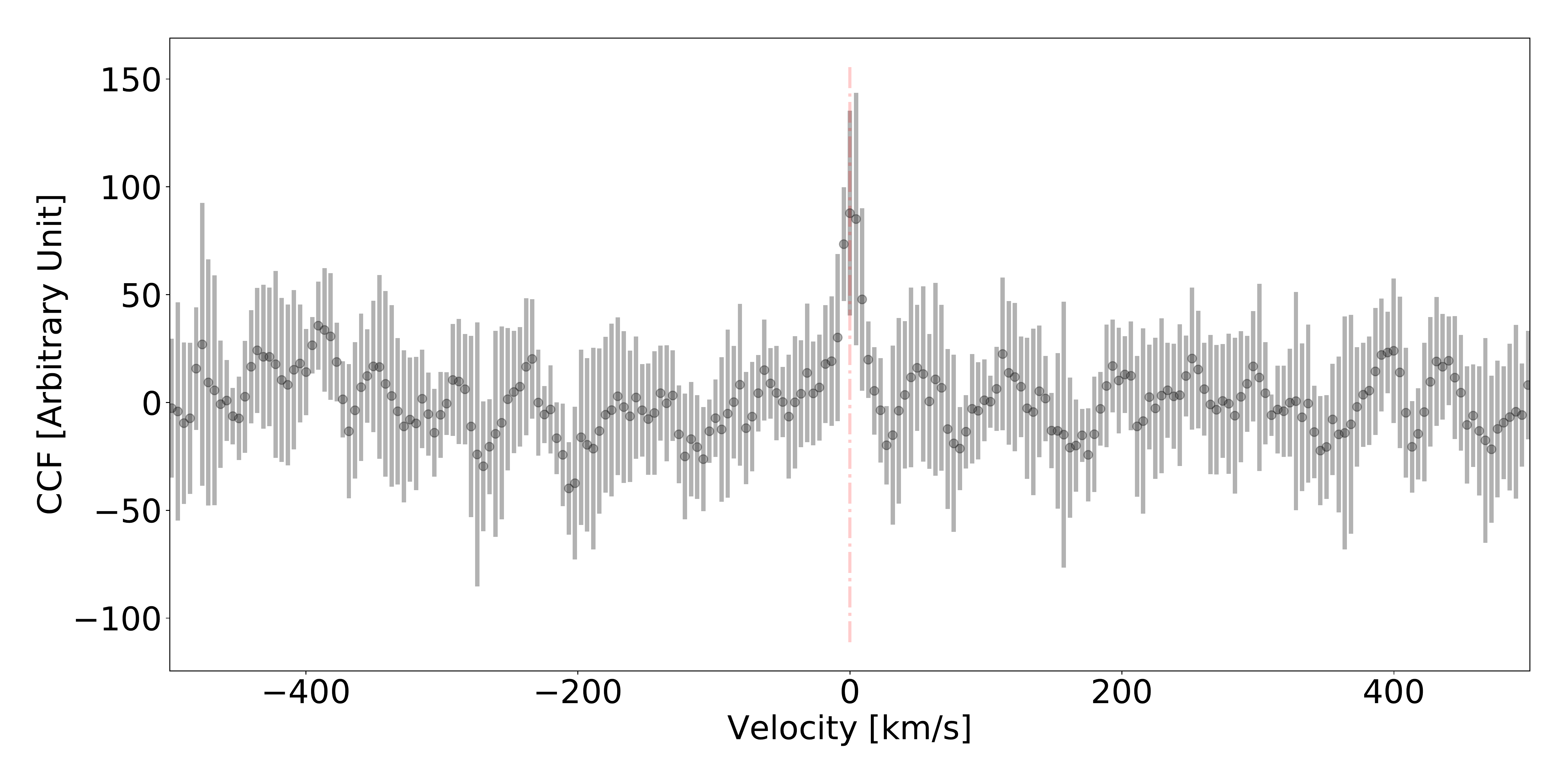}
\caption{Cross-correlation function for an injected planet signal (planet-star contrast = $2\times10^{-4}$). The injection can be detected at 4.9-$\sigma$. Red dashed line indicates stellar RV.    
\label{fig:ccf_error_injection}}
\end{figure*} 

We inject a planet signal that is $2\times10^{-4}$ times fainter than the central star (i.e., the planet-star contrast in $L$ band for HR 8799 c) at different angular distance to the star (0.909$^{\prime\prime}$, 0.963$^{\prime\prime}$, and 1.017$^{\prime\prime}$). We consider two scenarios: a CH$_4$-dominated atmosphere and an H$_2$O-dominated atmosphere. For the CH$_4$-dominated atmosphere, the injection signal is detected at 3.3-$\sigma$, 4.1-$\sigma$, and 4.9-$\sigma$ level (shown in Fig. \ref{fig:ccf_error_injection}), suggesting that $2\times10^{-4}$ sensitivity can be achieved with the reduced data. For H$_2$O-dominated atmosphere, the injection signal is detected at 6.3-$\sigma$, 5.5-$\sigma$, and 5.5-$\sigma$ level. The comparison between the two scenarios indicates that detection significance depends on the dominating opacity. When compared to the result of the 4.6-$\sigma$ detection of H$_2$O and the non-detection of CH$_4$ in real data, there are two implications: (1), HR 8799 c is likely to have a H$_2$O-dominated atmosphere; and (2), the template we use in cross-correlation is not a perfect match to the spectrum of HR 8799 c, resulting in a decrease of detection significance from $>$5.5-$\sigma$ to 4.6-$\sigma$. 

In addition, when compared to simulations in \S \ref{sec:simul} that account for all known noise sources (Table \ref{tab:simul_result}), detection significance reduces by a factor of 3.4. This implies that an additional noise source is not accounted for in the simulations in \S \ref{sec:simul}. The unknown noise could come from either the data reduction and analysis processes, or the noise introduced by the instrument and the detector. 

\subsection{Injecting Planet Signal In Raw Data}
\label{sec:injection_raw}
We also inject simulated planet signal into raw data. We create the injection signal the same way as described in \S \ref{sec:create_injection}. Injecting planet signal in raw data is different from planet injection in reduced data. The reduced data are rectified but raw data are curved and exhibits uneven order height (see \S \ref{sec:spec_rectification}). To simulate the curvature, we use the polynomial function that describes the stellar trace  to generate a curved planet spectrum. To account for the uneven order height, we use the measured expansion rate. These two treatments result in a planet trace that is in parallel with the stellar trace in the angular separation space. 


Similar to the injection experiment on reduced data, we inject a planet signal that is $2\times10^{-4}$ times fainter than the central star at different angular distance to the star (0.909$^{\prime\prime}$, 0.963$^{\prime\prime}$, and 1.017$^{\prime\prime}$). For CH$_4$-dominated atmosphere, the injection signal is detected at 2.7-$\sigma$, 4.1-$\sigma$, and 5.1-$\sigma$ level. For H$_2$O-dominated atmosphere, the injection signal is detected at 6.4-$\sigma$, 6.4-$\sigma$, and 5.9-$\sigma$ level. These numbers are comparable to the results for the reduced data injection, suggesting that the impact of data reduction and signal extraction is small. When combining results from the injection experiments for both the raw data and the reduced data, it is suggested that the noise is mainly from the unknown noise introduced by the instrument and the detector.

\section{Summary and Discussion}
\label{sec:summary}
We conduct high dispersion spectroscopy for HR 8799 and its planet c using Keck NIRSPEC in AO mode. We detect H$_2$O but not CH$_4$ in $L$ band. 

We conduct sensitivity analyses to investigate the detection threshold of our observations. The sensitivity analyses include (1), an end-to-end simulator accounting for a realistic noise budget and (2), planet signal injection experiments at different stages of data reduction and analysis. We conclude that the $L$-band observations have sufficient sensitivity to detect either CH$_4$ or H$_2$O, depending on which molecular species dominates $L$-band opacity.

\subsection{CH$_4$ Depletion vs. Incomplete Line List}
\label{sec:implications}


%

We detect H$_2$O but not CH$_4$ in $L$ band. This could be due to at least two reasons.  First, it is possible that there is a low abundance of CH$_4$ in the planet's atmosphere.  This would concur with the findings of \citet{Konopacky2013}.  Such a low abundance, given the planet's effective temperature, is at odds with predictions from equilibrium chemistry.  This is a now well-known phenomenon in field T dwarfs and should be expected in imaged planets \citep[e.g.][]{Marley2012}, caused by gas from the deeper, hotter atmosphere, being brought up to the visible atmosphere via convection.  A slow chemical conversion timescale for CO converting to (thermochemically favored) CH$_4$ leads to an atmospheric enriched in CO and depleted in CH$_4$ compared to equilibrium.  It has been suggested that at the low surface gravities of imaged planets, compared to brown dwarfs, this effect should be more pronounced \citep[e.g.,][]{Zahnle2014}.

Secondly, it is possible that CH$_4$ is indeed detectable (although weak) but the line positions from \citet{Yurchenko2014} are not modeled accurately enough to enable cross-correlation analysis.  To our knowledge there has not yet been a $L$-band cross-correlation detection of CH$_4$ in a brown dwarf, imaged planet, or transiting planet. The only $K$-band detection of CH$_4$ with the cross correlation technique has been made for HR 8799~b at R$\sim$~4000 by~\citet{Barman2011}. First-principles calculations of warm/hot CH$_4$ line positions are still developing and there is no yet suitable high-temperature high-resolution data for detailed and direct comparisons.

\subsection{Mixing Ratio of CH$_4$}

\begin{figure*}
\epsscale{1.1}
\plotone{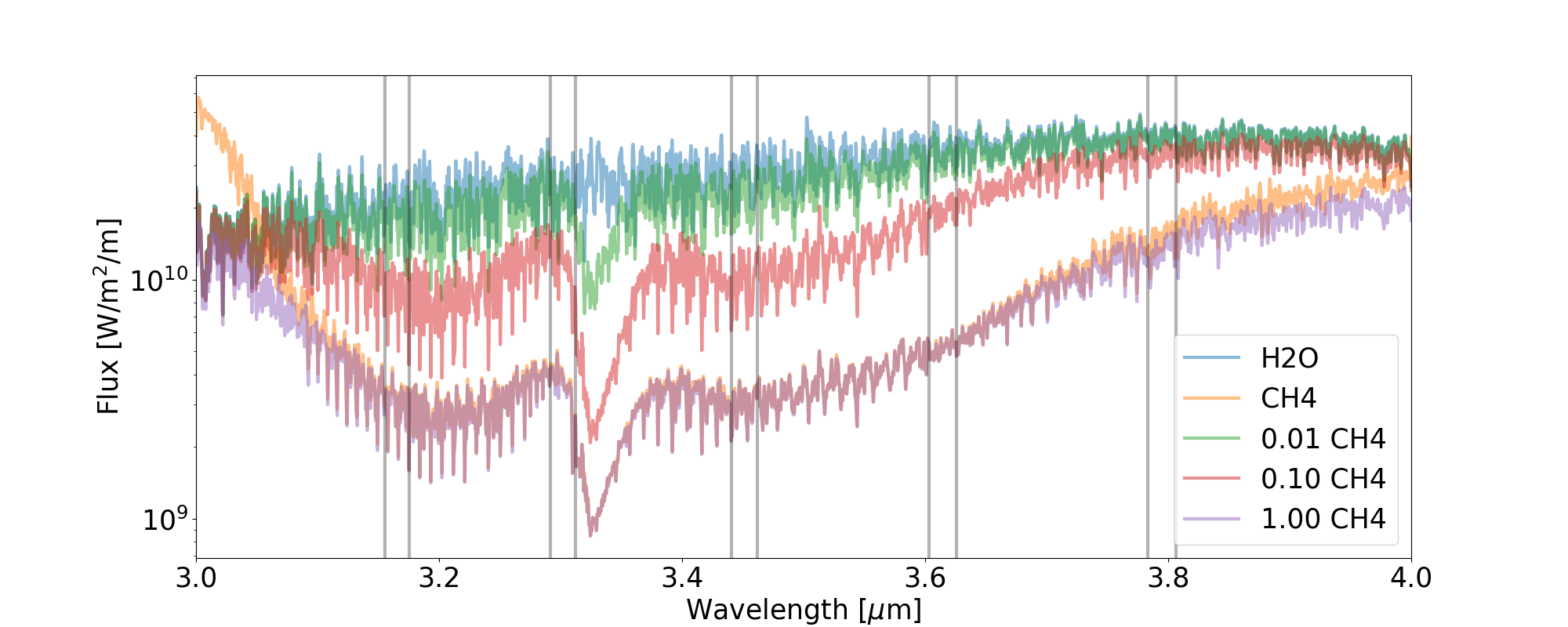}
\caption{Model spectra at different CH$_4$ mixing ratios. Spectral orders in $L$ band for NIRSPEC are marked by regions between grey lines.  
\label{fig:model_mixing_ratio}}
\end{figure*}

As previously discussed, we find the L-band spectrum to be dominated by H$_2$O lines instead of CH$_4$ lines.  We can investigate at what CH$_4$ mixing ratio H$_2$O lines start to dominate the spectrum.  We take the same atmosphere model but decrease the CH$_4$ mixing ratio with respect to the equilibrium abundance. Fig. \ref{fig:model_mixing_ratio} shows that H$_2$O lines start to appear at 10\% of CH$_4$ mixing ratio. At less than 10\% CH$_4$ mixing ratio, H$_2$O becomes the major opacity at wavelengths longer than 3.5 $\mu$m. However, CH$_4$ signal is clearly seen even when mixing ratio is at 1\%. This implies that the CH$_4$ mixing ratio could be much lower than 1\% of the equilibrium value.

{{In order to further examine the level at which CH$_4$ is depleted, we use a template in which CH$_4$ is depleted by 1000 times with respect to the chemical equilibrium value. The resulting CCF SNR is 4.4, lower than the CCF SNR by using a H$_2$O-only template (see \S \ref{sec:result}). We note that the difference may be due to random fluctuation in data processing procedure. However, if the difference is astrophysical in origin, then it indicates that a H$_2$O-only template matches better than the CH$_4$-depleted template. Therefore, CH$_4$ may be depleted by more than 1000 times with respect to the chemical equilibrium value.}}

\subsection{Modeling Uncertainty}
\label{sec:model_un}

A high fidelity template is the key to improve the sensitivity of high dispersion spectroscopy. No CCF peak would appear if the template is physically off the true spectrum. Results in Table \ref{tab:simul_result} suggest that a mismatch between an input spectrum and a template spectrum would reduce detection significance by at least a factor of 3 . In practice, we find it helpful to generate models with varying CH$_4$ mixing ratio. The variation mimics the effect of vertical mixing, which is largely unconstrained by observation at the moment. In general, varying atmospheric modeling parameters and exploring a large physically-motivated parameter space is necessary in detecting molecules and understanding the physical and chemical processes in exoplanet atmospheres. 




\subsection{The Prospect of Upgraded NIRSPEC}
We show in Table \ref{tab:simul_result} that planet c should be detected if accounting for realistic noise sources. However, CCF SNR is reduced in injection experiments. Planet signal must be lost at some stages of data recording, reduction and analysis. 

Conducting sensitivity analyses allows us to separate issues that cause planet signal loss. By comparing results from injections into the reduced data and the raw data, we conclude that the planet signal is not significantly weakened from raw to reduced data. This points to one stage where planet signal may be significantly weakened, i.e., from entering the telescope to the detector. 

{{It has been known that NIRSPEC detector has a correlated noise issue, most noticeable at long wavelengths ($L$ and $M$ band). However, this correlated noise has never been properly understood (private communication with Ian McLean, PI of NIRSPEC). We speculate here some possible sources for the observed correlated noise. First, there are repeated patterns every 8th column. This is likely to be caused by a particularly noisy output channel. Second, the periodical noise seen in Fig. \ref{fig:illus_L} has a periodicity of $\sim$16 pixels. This is possibly due to some temporal noise of output channels. During the readout, every 8 columns are read by 8 output channels for each quadrant. It is possible that correlated noise appears when readout frequency beats with the frequency of the temporal noise. Third, if the temporal noise has a much lower frequency than the pixel rate, then the noise would make certain columns to have higher or lower values than the average value, causing the correlated noise that we see along columns. Finally, if the correlated noise comes from the detector itself, then the noise may be from the stage of transistor and/or multiplexer. }}

The situation can be improved by the detector upgrade for NIRSPEC that is currently ongoing~\citep{Martin2014}. The upgraded NIRSPEC will be available starting in 2018B. The improved sensitivity will allow detection of CH$_4$ in $L$ band based on our simulation. When combined with KPIC~\citep{Mawet2017}, HDC will be demonstrated at Keck and the combination will pave the way for future HDC instruments at giant segmented mirror telescopes (GSMT).  

\noindent{\it Acknowledgements} 
We thank the anonymous referee for his/her constructive comments and suggestions that greatly improve the manuscript. We acknowledge Rowan Swain for carefully proofreading the manuscript. We thank Jason Wang and Olivier Wertz for providing precise astrometric predictions for HR 8799 c. We thank Roger Smith and the NIRSPEC team especially Ian McLean, Mike Fitzgerald, and Emily Martin for valuable input on the correlated noise of ALADDIN InSb detectors. We thank Geoff Blake and his group for offering advice for NIRSPEC $L$-band observation and sharing the PyNIRSPEC data reduction pipeline. The data presented herein were obtained at the W. M. Keck Observatory, which is operated as a scientific partnership among the California Institute of Technology, the University of California and the National Aeronautics and Space Administration. The Observatory was made possible by the generous financial support of the W. M. Keck Foundation. The authors wish to recognize and acknowledge the very significant cultural role and reverence that the summit of Maunakea has always had within the indigenous Hawaiian community.  We are most fortunate to have the opportunity to conduct observations from this mountain.

\bibliography{mybib_JW_DF_PH5}

\appendix
\section{A: Rectification and Wavelength Calibration}
\label{app:dr_details}

\subsection{Rectifying Spectral Order}
\label{sec:spec_rectification}
Rectifying spectral order for HR 8799 c data required special care because (1), two spectra needed to be extracted, one for the star and one for the planet; and (2), the planet spectrum was unseen. 

The situation was complicated by the shape of each spectral order. Along the dispersion direction, the trace of the spectrum curled smoothly. Along the slit direction, the height of each order differed at one end of the spectrum from the other end. Spectral orders at the blue end were 10 pixels higher than spectral orders at the red end (130 pixels vs. 120 pixels) for the $L$ band data. 

Without correcting for the uneven order height, straightening the HR 8799 stellar spectrum would result in a tilted HR 8799 c planet spectrum. Consequently, unwanted noise would be introduced when extracting the HR 8799 c spectrum. Therefore, we corrected for the order height distortion by expanding/shrinking each spectral column such that order heights were the same throughout all spectral channels. After the correction, order height was 130 pixels. 

We then proceeded to find the trace of the stellar spectrum and rectified the spectrum. This process also rectified the planet spectrum. We calculated the flux centroid offset along the slit direction using the middle part of the spectrum as a reference. This was done by cross-correlating flux of each column with the flux of the middle column. We fitted the centroid offset as a function of pixel with a third-ordered polynomial function. We then used the function to shift the flux so that the stellar spectrum was straight and oriented horizontally.  

Raw data in $L$ band exhibited strong sky emission lines (see Fig. \ref{fig:HR8799c_L_raw_order}). These lines did not align with the vertical direction on detector. This misalignment is because the slit was slightly tilted with respect to the detector column direction. Thus, the same column on the detector did not always have the same wavelength solution, i.e., there is a slight wavelength shift from row to row. The tilted slit caused a problem in removing sky background emission: subtracting the median flux of each column would result in residuals on the top and bottom parts of the spectrum. 

Therefore, we needed to correct for the ``tilted slit". The process is similar to straightening the stellar spectrum. We cross-correlated each row (excluding rows with the stellar spectrum trace) and found the relative shift. The shift as a function of row number was fitted by a third-ordered polynomial. Flux of each row was then shifted so that the orientation of the slit was aligned with the vertical direction of the detector. 

We then removed sky background emission by subtracting the median flux of each column. The rectified sky-subtracted 2d spectra are shown on the right of Fig. \ref{fig:HR8799c_L_raw_order}. 

\subsection{Wavelength Calibration}
\label{sec:wav_sol}
We used telluric absorption lines in the stellar spectrum as a reference for wavelength calibration. Through comparison with synthetic telluric absorption spectra generated by the HITRAN database~\citep{Rothman2009}, we identified lines and recorded their corresponding pixel {{locations}} and wavelengths. We then fitted wavelengths of identified lines as a function of pixel with a 4th-order polynomial function. The typical fitting residual was $1\times10^{-5}-2\times10^{-5}$ $\mu$m. This residual corresponds to 1-2 km$\cdot\rm{s}^{-1}$ in RV shift. Because of prominent noise patterns on the left side of detector (Fig. \ref{fig:HR8799c_L_raw_order}), we used only the right half of the detector for the following data reduction and analyses. 


There are 5 spectral orders in $L$-band data. The wavelength ranges for the 5 orders were 3.783-3.806 $\mu$m (order 1), 3.604-3.626 $\mu$m (order 2), 3.441-3.462 $\mu$m (order 3), 3.292-3.312 $\mu$m (order 4), and 3.156-3.175 $\mu$m (order 5).

\clearpage

\newpage

\begin{deluxetable}{lcccccccc}
\tablewidth{0pt}
\tablecaption{Summary of $L$ band observations for HR 8799 c .\label{tab:obs_summary}}
\tablehead{
\colhead{\textbf{Date}} &
\colhead{\textbf{Starting}$^{\ast}$} &
\colhead{\textbf{Ending}$^{\ast}$} &
\colhead{\textbf{Duration}} &
\colhead{\textbf{Seeing}} &
\colhead{\textbf{Separation}} &
\colhead{\textbf{PA}} &
\colhead{\textbf{Star$^{\ast\ast}$}} &
\colhead{\textbf{Sky$^{\ast\ast}$}}\\
\colhead{\textbf{UT}} &
\colhead{\textbf{hh:mm}} &
\colhead{\textbf{hh:mm}} &
\colhead{\textbf{hh:mm}} &
\colhead{\textbf{arcsec}} &
\colhead{\textbf{mas}} &
\colhead{\textbf{degree}} &
\colhead{\textbf{ADU}} &
\colhead{\textbf{ADU}} 
}

\startdata

2016 Aug 12 & 10:54 & 15:14 & 4:20 & 0.60$^{\prime\prime}$ & 944.4$\pm$0.9 & 329.8$\pm$0.1 & 7969 & 1015-1859 \\
2016 Aug 13 & 10:50 & 15:18 & 4:28 & 0.60$^{\prime\prime}$ & 944.4$\pm$0.9 & 329.8$\pm$0.1 & 8438 & 1066-1588 \\
2017 Jul 06 & 10:34 & 15:04 & 4:30 & 0.80$^{\prime\prime}$ & 942.0$\pm$1.3 & 331.4$\pm$0.1 & 7273 & 970-1702 \\
2017 Nov 06 & 05:22 & 10:14 & 4:52 & 0.54$^{\prime\prime}$ & 941.2$\pm$1.5 & 331.9$\pm$0.1 & 4902 & 1131-2609 \\

\enddata

\tablecomments{$\ast$: in UT. $\ast\ast$: per pixel at 3.8 $\mu$m.}

\end{deluxetable}
 
\newpage

\begin{deluxetable}{ccc}
\tablewidth{0pt}
\tablecaption{Telescope and instrument parameters for simulated observations of HR 8799 c .\label{tab:telescope_instrument}}
\tablehead{
\colhead{\textbf{Parameter}} &
\colhead{\textbf{Value}} &
\colhead{\textbf{Unit}} \\
}

\startdata

Telescope aperture & 10.0 & m \\
Spectral resolution & 15,000 & \nodata \\
Pixels per resolution element & 5.0 & \nodata \\
$L$ band spectral range & [3.783, 3.806], [3.604, 3.626], [3.441, 3.462], [3.292, 3.312], [3.156, 3.175] & $\mu$m \\
Exposure time & 41580 & second \\
Slit width & 1.0 & $\lambda$/D \\
Wavefront correction residual$^\ast$ & 260 & nm \\
Star light reduction at planet position & $10^{-3}$ & \nodata \\
Telescope+instrument throughput$^{\ast\ast}$ & 1.4\% & \nodata \\
Readout noise & 23 & e$^{-}$ \\
Number of Readouts & 460 & \nodata \\
Dark current & 0.8 & e$^{-}$ s$^{-1}$

\enddata

\tablecomments{$\ast$: Private communication with Peter Wizinowich. $\ast\ast$: The throughput is calculated so that simulated stellar continuum level is the same as the observed stellar continuum level. }

\end{deluxetable}
 
\newpage

\begin{deluxetable}{cccc}
\tablewidth{0pt}
\tablecaption{HR 8799 and planet c.\label{tab:HR8799c}}
\tablehead{
\colhead{\textbf{Parameter}} &
\colhead{\textbf{Value}} &
\colhead{\textbf{Unit}} &
\colhead{\textbf{References}} \\
}

\startdata

\multicolumn{4}{l}{\textbf{Star}} \\
Effective temperature (T$_{\rm{eff}}$) & 7193 & K & ~\citet{Baines2012} \\
Surface gravity ($\log g$) & 4.03 & cgs & ~\citet{Baines2012} \\
Distance & 39.40 & pc & ~\citet{vanLeeuwen2007} \\
V$\sin i$ & 37.5 & km s$^{-1}$ & ~\citet{Kaye1998} \\
Radial velocity & 12.4 & km s$^{-1}$  & ~\citet{Gontcharov2006} \\
\multicolumn{4}{l}{\textbf{Planet}} \\
Effective temperature (T$_{\rm{eff}}$) & 1100-1350 & K & ~\citet{Bonnefoy2016} \\
Surface gravity ($\log g$) & 3.5-3.9 & cgs & ~\citet{Bonnefoy2016} \\
Metallicity ([M/H]) & 0.0-0.5 & dex & ~\citet{Bonnefoy2016} \\
Semi-major axis ($a$) & 42.8 & AU & ~\citet{Zurlo2016} \\
Angular separation$^{\ast}$ & 0.941-0.944 & arcsec & ~\citet{Wertz2017}, ~\citet{Wang2016} \\
Planet/star contrast in $L$ & $2\times10^{-4}$ & \nodata & ~\citet{Currie2014} \\
\enddata


\end{deluxetable}
 
\newpage

\begin{deluxetable}{ccc}
\tablewidth{0pt}
\tablecaption{Detection Significance Based on Simulated Observations for HR 8799 c.\label{tab:simul_result}}
\tablehead{
\colhead{\textbf{Scenario:}} &
\colhead{\textbf{Mismatch}} &
\colhead{\textbf{Match}} \\
\colhead{\textbf{Input:}} &
\colhead{\textbf{Fortney 1\% CH$_4$}} &
\colhead{\textbf{Fortney 100\% CH$_4$}} \\
\colhead{\textbf{Template:}} &
\colhead{\textbf{Fortney}} &
\colhead{\textbf{Fortney}} \\
\colhead{\textbf{Order$^{\ast}$}} &
\colhead{\textbf{all$^{\ast\ast}$,CH$_4$,H$_2$O}} &
\colhead{\textbf{all,CH$_4$,H$_2$O}} \\
}

\startdata
1 & 1.63, -0.16, 4.41 & 5.05, 4.71, 0.97 \\
2 & 1.48, 0.89, 7.76 & 2.55, 2.27, 0.43 \\
3 & 3.54, 2.77, 7.10 & 2.68, 3.00, 0.66 \\
4 & 0.43, -0.64, 5.73 & 2.92, 2.78, 0.26 \\
5 & 2.65, 2.11, 9.05 & 2.95, 2.96, 0.46 \\
All$^{\ast\ast\ast}$ & 4.96, 3.59, 15.65 & 7.51, 7.23, 1.34 \\

\enddata

\tablecomments{$^{\ast}$: See \S \ref{sec:wav_sol} for wavelength range for each order. $^{\ast\ast}$: use Fortney 100\% CH$_4$ model as a template. $^{\ast\ast\ast}$: Summation of quadrature of detection significance for all orders with higher than zero detection significance.}

\end{deluxetable}
 
\end{document}